\documentclass[twocolumn]{aastex61}
\usepackage{amsmath}
\begin{document}

\title{Deep recurrent neural networks for supernovae classification}

\author{Tom Charnock} \email{tom.charnock@nottingham.ac.uk}
\affiliation{School of Physics \& Astronomy\\
University of Nottingham,
Nottingham, NG7 2RD, England}

\author{Adam Moss} \email{adam.moss@nottingham.ac.uk}
\affiliation{School of Physics \& Astronomy\\
University of Nottingham,
Nottingham, NG7 2RD, England}

\date{\today}

\begin{abstract}
   We apply deep recurrent neural networks, which are capable of learning complex sequential information, to classify supernovae\footnote{Code available at \href{https://github.com/adammoss/supernovae}{https://github.com/adammoss/supernovae}}. The observational time and filter fluxes are used as inputs to the network, but since the inputs are agnostic additional data such as host galaxy information can also be included. Using the Supernovae Photometric Classification Challenge (SPCC) data, we find that deep networks are capable of learning about light curves, however the performance of the network is highly sensitive to the amount of training data.  For a training size of 50\% of the representational SPCC dataset (around $10^4$ supernovae) we obtain a type-Ia vs. non-type-Ia classification accuracy of 94.7\%, an area under the Receiver Operating Characteristic curve AUC of 0.986 and a SPCC figure-of-merit $F_1=0.64$. When using only the data for the early-epoch challenge defined by the SPCC we achieve a classification accuracy of 93.1\%, AUC of 0.977 and $F_1=0.58$, results almost as good as with the whole light-curve. By employing bidirectional neural networks we can acquire impressive classification results between supernovae types -I,~-II and~-III at an accuracy of 90.4\% and AUC of 0.974. We also apply a pre-trained model to obtain classification probabilities as a function of time, and show it can give early indications of supernovae type. Our method is competitive with existing algorithms and has applications for future large-scale photometric surveys. 
   \end{abstract}

\section{Introduction}

Future large, wide-field photometric surveys such as the Large Synoptic	Survey Telescope (LSST) will produce a vast amount of data, covering a large fraction of the sky every few nights. The amount of data produced lends itself to new analysis methods which can learn abstract representations of complex data. Deep learning is a powerful method for gaining multiple levels of abstraction, and has recently produced state-of-the-art results in tasks such as image classification and natural language processing (see~\cite{0483bd9444a348c8b59d54a190839ec9} for an excellent overview of deep learning and refs. within for more details).

There are many applications of deep learning for large photometric surveys, such as: (1) the measurement of galaxy shapes from images; (2) automated strong lens identification from multi-band images; (3) automated classification of supernovae; (4) galaxy cluster identification. In this paper we will focus on supernovae classification using deep recurrent neural networks. The LSST, for example, is expected to find over $10^7$ supernova~\cite{2009arXiv0912.0201L}. However,  it is estimated that only 5000 to 10,000\footnote{Although these numbers are not guaranteed.} will be spectroscopically confirmed  by follow up surveys~\cite{2013arXiv1311.2496M}, so classification methods need to be developed  for photometry.  All previous approaches to automated classification~\cite{Newling:2010bp, Karpenka:2012pm, Lochner:2016hbn} have first extracted features from supernovae light curves before using machine learning algorithms.  One of the advantages of deep learning is replacing this feature extraction. 

In this work we will use {\em supervised} deep learning. During training, the machine is given inputs and produces a set of output predictions.  It is also given the correct set of outputs. An objective loss function then measures the error between the predicted and target outputs, and the machine updates its adjustable parameters to reduce the error. It can then make predictions for unknown outputs.

\begin{figure}
\centering
\includegraphics[width=83mm, angle=0]{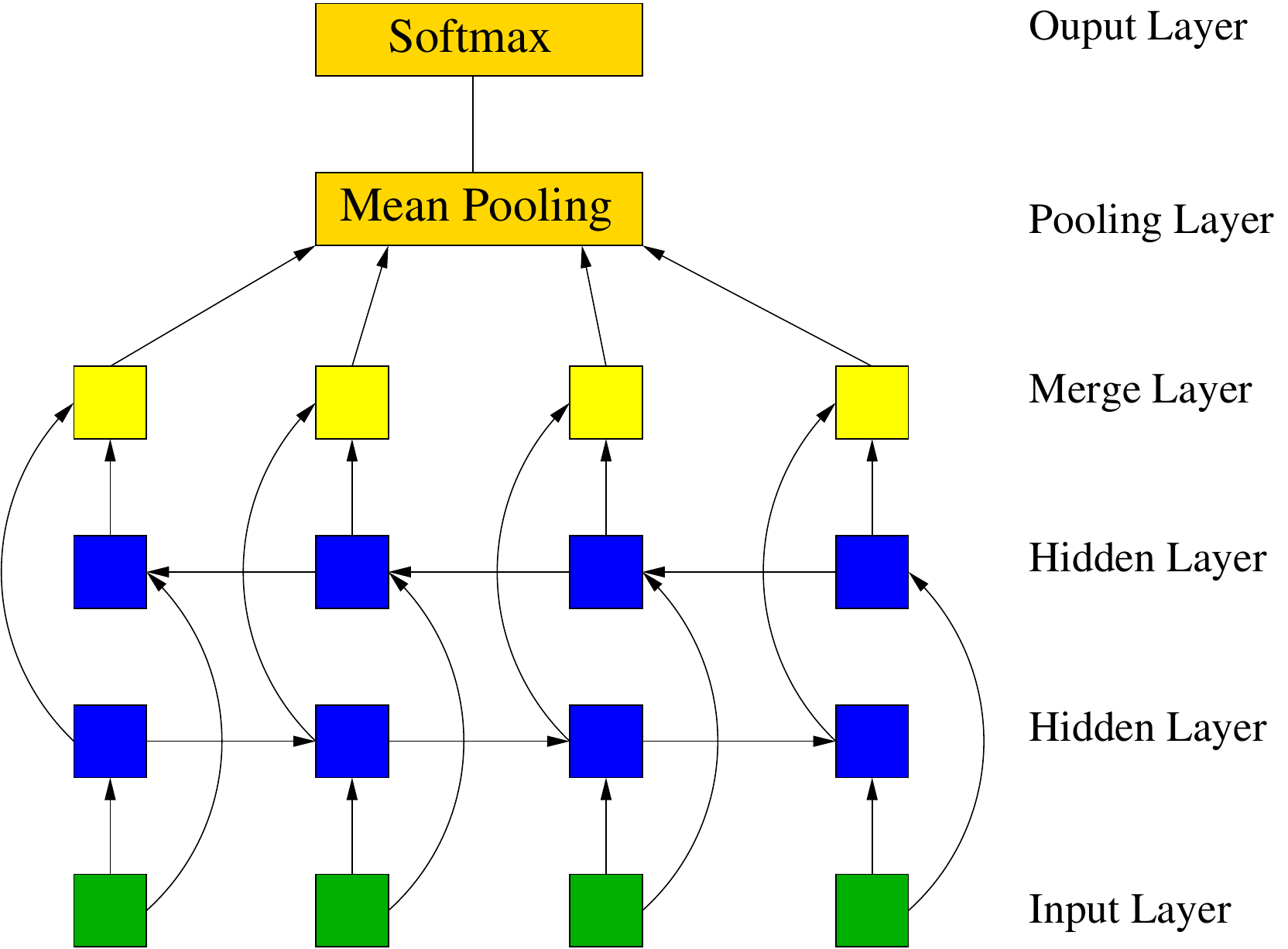}
\caption{\label{fig:network} Bidirectional recurrent neural network for sequence classification. The input vectors at each sequential step are fed into a pair of bidirectional hidden layers, which can propagate information forwards and backwards. These are then merged to obtain a consensus view of the network, and finally a softmax layer computes classification probabilities. 
 }
\end{figure}

{\em Recurrent neural networks} (RNNs) are a class of artificial neural network  that can learn about sequential data (for an extremely comprehensive overview see~\cite{medsker1999recurrent}). They are commonly used for tasks such speech recognition and language translation, but have several possible applications in astronomy  and cosmology for processing temporal or spatial sequential data. RNNs have several properties which makes them suitable for sequential information. The inputs to the network are flexible, and they are able to recognise patterns with noisy data (for example the context of a word in a sentence relative to others can vary, or a time stream can contain instrument noise). 

The main problem with vanilla RNNs  is that they are unable to store long-term information, so inputs at the end of a sequence have no knowledge of inputs at the start. This is a problem if the data has long-term correlations. Several types of RNNs have been proposed to solve this problem, including {\em Long Short-Term Memory} (LSTM) units~\cite{LSTM} and {\em Gated Recurrent Units} (GRU)~\cite{2014arXiv1412.3555C}. These are similar in concept, in that information is able to flow through the network via a gating mechanism. Another problem with RNNs is that information can only flow in one direction. In {\em bidirectional} RNNs information is able to pass both forwards and backwards. Bidirectional LSTM networks have been shown to be particularly powerful where sequential data is accompanied by a set of discrete labels. 

The architecture of a typical bidirectional RNN for sequence labelling is shown in Fig.~\ref{fig:network}, where the squares represent {\rm neurons}. In this case the inputs, which are vectors at each sequential step, are connected to two hidden RNN layers, either vanilla RNN or memory units.  Each hidden layer contains a number of hidden units (capable of storing information), and in each layer information flows either forwards or backwards,  but no information passes between the two directions. Several hidden layers can be stacked to form {\em deep} neural networks. Deep networks are capable of learning higher-level temporal or spatial representations, and complex relationships between the inputs and outputs.

The output from the final set of hidden layers in each direction is merged at each sequential step, and mean pooled (averaged) over all steps to obtain a consensus view of the network\footnote{We find that obtaining a consensus view improves the performance of the network.}. Finally, the mean output is fed to a {\em softmax} layer, taking an input vector ${\bf z}$ and returning normalised, exponentiated outputs for each class label $i$, $\exp(z_i) / \sum_{i} \exp(z_i)$, i.e. a vector of probabilities.

 Each neuron is connected to another by a weight matrix, and the optimal weights are found by back-propagating the errors from a {\em loss function} of the output layer. For classification problems, this is typically the categorical cross-entropy between predictions and targets, defined as 
\begin{equation}
L= -\sum_{i,j} t_{i,j} \log \left( p_{i,j} \right)
\end{equation}
where $i,j$ run over the class labels, $t_{i,j}$ are the targets for each class (either 0 or 1) and $p_{i,j}$ are the predicted probabilities. Back-propagation takes the derivative of the loss with respect to the weights $W$ of the output layer, $\partial L/\partial W$, and uses the chain rule to update the weights in the network.

\section{Example Data}

\begin{figure}
\centering
\includegraphics[width=84mm, angle=0]{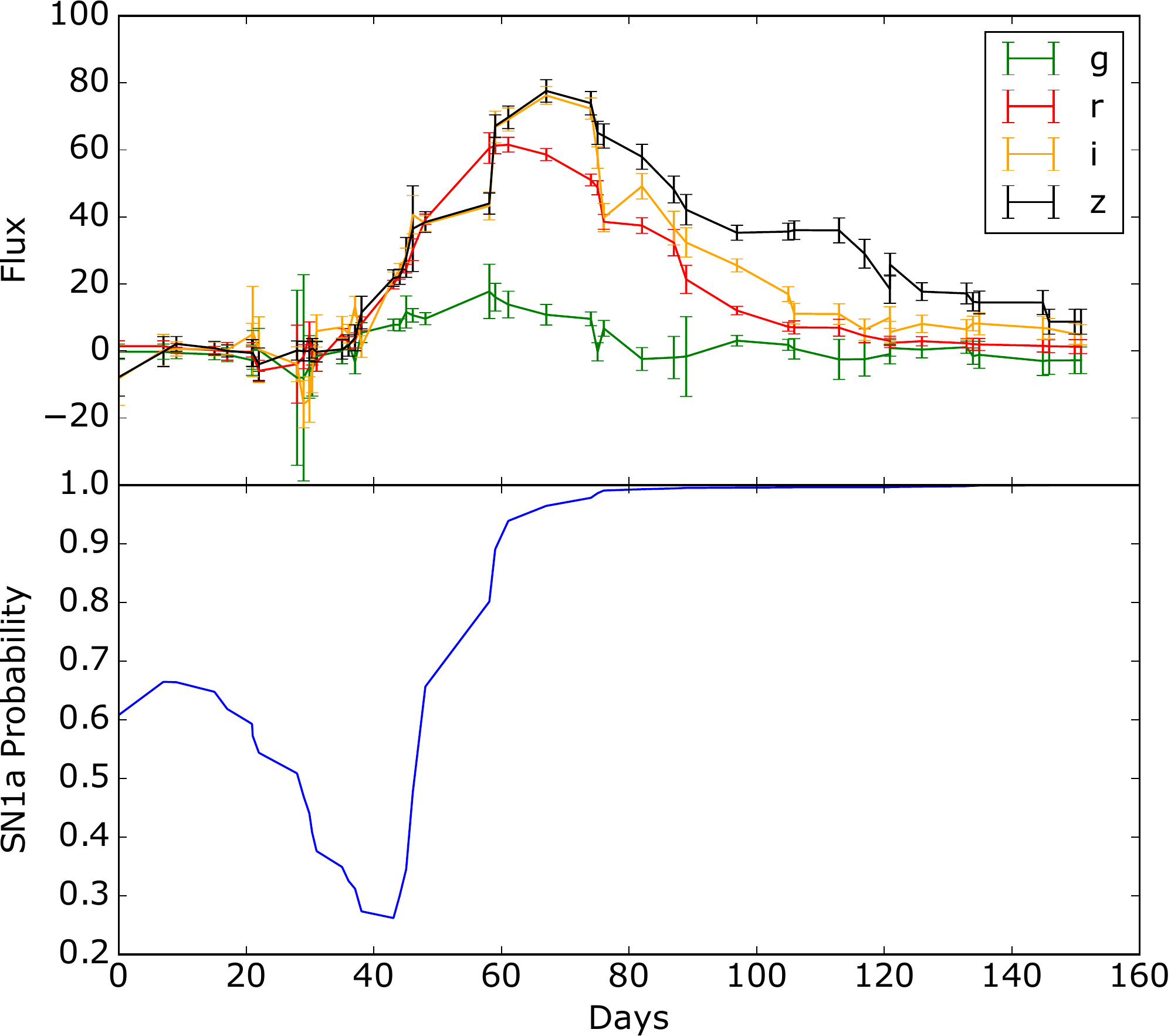}
\caption{\label{fig:lightcurve} (Top) Example light curve in the 4 $g,r,i,z$ bands for SN ID 551675 (a type-Ia) in the Supernovae Photometric Classification Challenge data~\cite{Kessler:2010wk}. The data has been processed using augmentation so there is a $g,r,i,z$ value at each sequential step. (Bottom) Type-Ia probability as a function of time from a 2 layer LSTM model, trained with around $10^4$ supernovae and SN 551675 excluded. The final probability gives $99.5\%$ confidence that the supernovae is of type-Ia. 
 }
\end{figure}

In this paper we will consider data from the Supernovae Photometric Classification Challenge (SPCC)~\cite{Kessler:2010wk,Kessler:2010qj}, consisting of 21319 simulated supernova light curves.  Each supernovae sample consists of a time series of flux measurements, with errors, in the $g,r,i,z$ bands (one band for each timestep), along with the position on the sky and dust extinction. An example set of light curves is shown in Fig.~\ref{fig:lightcurve}. 

Due to the format of the input data, we first do a small amount of data processing to obtain values of the $g,r,i,z$ fluxes and errors at each sequential step.  We assume the time sequence begins at day 0 for each supernovae, rather than counting days forwards and backwards from the maxima of the light curve. For observations less than $\sim1$ hour apart, we group the $g,r,i,z$ values into a single vector, ensuring there is at most one filter-type in each group. If there is more than one filter-type, we further subdivide the group using a finer time interval.  The group time is the mean of the times of each observation, which is reasonable as the time intervals are small compared to the characteristic time of the light curve. 

\begin{figure}
\centering
\includegraphics[width=84mm, angle=0]{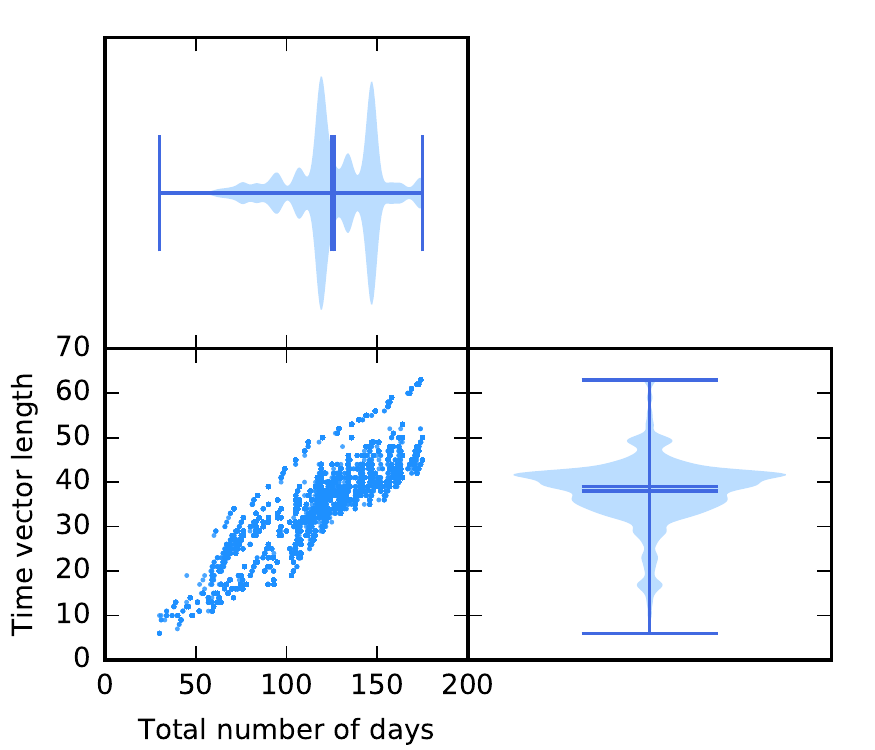}
\caption{\label{fig:violinplot}. (Top) Distribution of the total number of days for each light-curve with the minimum, maximum, mean and median values indicated. (Bottom right) Distribution of the number of elements in the grouped time vector with the minimum, maximum, mean and median values indicated. (Bottom left) The trend showing that more days in the light-curve result in longer group time vectors.
 }
\end{figure}

In Fig.~\ref{fig:violinplot} we show how the length of the grouped-time data vector is related to the duration of the light-curve. The bottom left subplot shows that more total number of day since the beginning of observation of the light-curve results in a greater number of grouped time elements in the vector. The upper subplot shows the distribution of observation lengths in the SPCC data varies significantly with two distinct peaks. These are grouped into an average of 40-element data vectors as can be seen in the bottom right subplot.

Observations are of the form in Table~\ref{tab:augment}, where any missing values are denoted by a dash. In order to impute the missing value of $i$, we use {\em data augmentation} and randomly select a value between $i_1$ and $i_3$. We make 5 random augmentations of all missing data, thereby increasing the size of the dataset fivefold. We can test the importance of this by training each augmentation separately and comparing the change in accuracy, which we find is $\sim 1\%$. Training with multiple augmentations at once gives the best performance since the network learns to ignore random-filled values.

\begin{table}
\begin{center}
\begin{tabular}{|c|c|c|c|c|}\hline  
 Time&g&r&i&z\\\hline\hline 
$t_1$&$g_1$&$r_1$&$i_1$&$z_1$\\\hline 
$t_2$&$g_2$&$r_2$&$-$&$z_2$\\\hline 
$t_3$&$g_3$&$r_3$&$i_3$& $z_3$\\\hline 
 \end{tabular}
\end{center}
\caption{\label{tab:augment} Data augmentation of missing observations. The missing data is replaced randomly by a value between $i_1$ and $i_3$.}
\end{table}%

The data comes in two types, those with and those without the host galaxy photometric redshift. Each dataset is split into a training and test set, with the training set containing a spectroscopically confirmed supernovae type and redshift. It is important that augmented data with the same supernovae ID go into either the training {\em or} test set otherwise they will not be independent. The original SPCC data consisted of 1103 training samples. The answer keys were subsequently made available for the test set~\cite{Kessler:2010qj}. 

The input vector to each sequential step consists of: time in days since the first observation; flux in each of the 4 bands; flux errors in each of the 4 bands; RA and Dec; dust extinction; and host photo-z if relevant. Whilst we do not expect some of these variables to impact the classifier accuracy, we do not attempt any feature engineering and leave it to the network to decide if they are relevant. 

RNNs typically perform better with more training data, so we train using the SPCC test set with answer keys (which is a non-biased representational dataset\footnote{The original SPCC training set was non-representational.}), and select a random fraction to act as the training set. We consider 1103 supernovae (a training fraction of 0.052), the same size as the original challenge, and fractions of 0.25 and 0.5 (around 5000 and $10^4$ supernovae respectively), nearly an order of magnitude larger, and closer to the number likely to be followed up for the LSST. The training performance of RNNs is also improved if the data is processed in mini-batches. In order to do this the input data must be of the same length, so we set the sequence length to be the maximum length over all supernovae observations, and prepend the input with padding. In training the network we ensure the padding is ignored by masking the padded input. 
 
The times of the observations in the light-curve are irregularly spaced and whilst this may not be optimal for the network we find that it is better to use the data padded at the end of the sequence than to place observations at similar times in similar sequence positions. There may even be hidden connections between the clustering of observation times and supernovae type, although it is hard to test for this.

The goal of the classifier is to determine the supernovae type in the test set. We consider two problems, (1) to categorise two classes (type-Ia vs. non-type-Ia), and (2) to categorise three classes (supernovae types-1, -2 and -3). We denote these as  `SN1a' and  `123' respectively. We also attempt the first two problems using only the first six observations with $S/N>4$ and the data taken on the night of the sixth observation as described in~\cite{Kessler:2010wk}.

Several metrics are used to assess the classifier. The simplest is the accuracy, defined as the ratio between the number of correct predictions and total number of predictions. With two classes, a random classifier would have an accuracy of 0.5, and with three classes, an accuracy of 1/3.

Next are a variety of metrics coming from the {\em confusion matrix} of predictions.  For binary classification problems, the confusion matrix splits predictions into true positives~(TP), false positives~(FP), false negatives~(FN), and true negatives~(TN). We consider the purity and completeness of the classifier. These are defined as 
\begin{equation}
{\rm Purity} = \frac{{\rm TP}}{{\rm TP}+{\rm FP}}\,, \quad {\rm Completeness} = \frac{{\rm TP}}{{\rm TP}+{\rm FN}}\,.
\end{equation}
We evaluate these for each class separately vs. `the rest' (e.g. type-Ia vs. non-type-Ia). The SPCC also defined the $F_1$ figure-of-merit for the SN1a classification problem. This is 
\begin{equation}
F_1 = \frac{1}{{\rm TP}+{\rm FN}} \frac{{\rm TP}^2}{{\rm TP}+3 \times {\rm FP}}\,,
\end{equation}
so incorrectly classifying a non-type-Ia supernovae as a type-Ia is penalised more heavily.

Finally, we calculate the Area-Under-the-Curve (AUC). The AUC is the area under the curve of the TP rate vs. FP rate, as the threshold probability for classification is increased from 0 to 1. A perfect classifier has an AUC of 1, and a random classifier 0.5. For multi-class problems, we calculate the AUC for each class vs. the rest, and take an unweighted average to give the final AUC score. 

\section{Network Architecture}

We consider several combinations of the network architecture. For the RNN type in the hidden layers, we test both vanilla RNN and long-term memory (LSTM and GRU) units. We also consider unidirectional and bidirectional networks. For unidirectional networks we fix the direction to be forwards. For bidirectional networks, the number of hidden units in each RNN layer is equal in the forward and backward directions. 

We also test stacking two sets of layers to form a deep network.  In the unidirectional case we stack two hidden layers. In the bidirectional case the two stacks consists  of a pair  of forwards and backwards layers.  We denote the number of hidden units in a network with a  single stack by $[h_1]$, and the number of hidden layers in a two stack model by $[h_1, h_2]$. We vary the number of hidden units, testing $h=[4],[8],[16],[32],[4,4],[8,8],[16,16]$ and~$[32,32]$. We do not go beyond a stack of two layers due to the limited size of the dataset. 

For each network we perform 5 randomised runs over the training data to obtain the classifier metrics. The loss function is the categorical cross-entropy between the predictions and test data. The network weights ware trained using back-propagation with the {\ttfamily Adam} updater~\cite{2014arXiv1412.6980K}. Mini-batches containing 10 samples\footnote{If training with a GPU larger mini-batches are recommended to make use of the GPU cores.} were used throughout, and each model was trained for 200 {\em epochs}, where each epoch is a full pass over the training data.

\section{Results}

\begin{figure}
\centering
\includegraphics[width=84mm, angle=0]{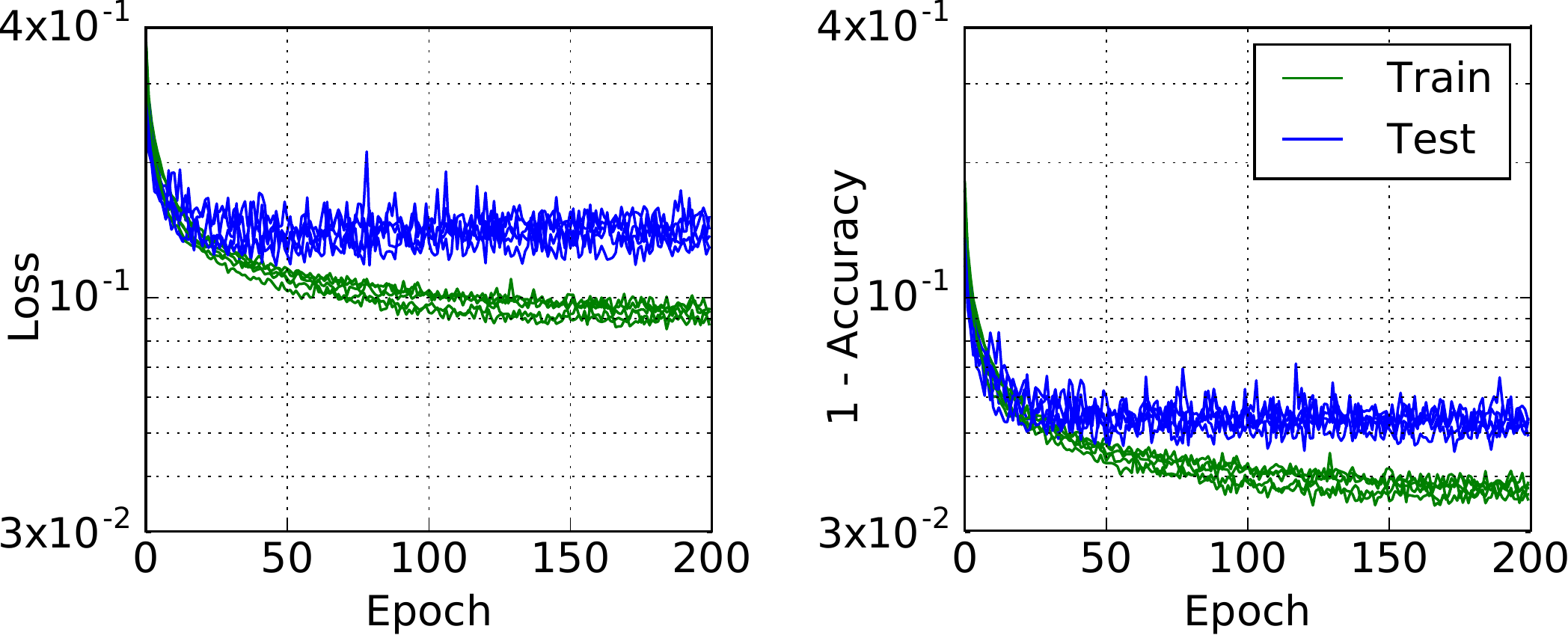}
\caption{\label{fig:loss} (Left) Training loss (green) vs. test loss (blue) for a unidirectional 2 layer LSTM network with 16 hidden units in each layer. (Right) Training accuracy (green) vs. test accuracy (blue) for the same network.  }
\end{figure}

A dataset of 21319 is relatively small by deep learning standards. Furthermore, the `feature space' of supernovae light curves is significantly smaller than, say, using RNNs to learn about language. We therefore need to be careful about {\em over-fitting}. Over-fitting arises when the network learns about relations between the inputs and outputs of the training data, which do not exist in the test data. It can typically be detected by comparing the loss of the training and test data. If the loss of training data continues to decrease, but the loss of the test data increases, this is a sure sign of over-fitting. If no sign of over-fitting is observed, the network is not usually complex enough to fully learn the relationship between inputs and outputs (called {\em under-fitting}).  

For a training fraction of 0.5, we found the best architecture was a deep 2-layer network with unidirectional LSTM units. Bidirectional units did not significantly improve the test accuracy and made the network more difficult to train. There was a marked improvement in test accuracy using 16 hidden units in each layer rather than 8, but too much over-fitting occurred using 32 hidden units. Over-fitting was still an issue for 16 hidden units,  but a technique called {\em dropout}~\cite{JMLR:v15:srivastava14a} could regularise this. Dropout sets a random fraction of connections to 0 at each update during training {\em only}, preventing the units from adapting too much. We apply dropout only to non-recurrent connections after each hidden layer. 

In Fig.~\ref{fig:loss} we show the training and tests losses for such a network, with a dropout of $0.5$, applied to type-Ia vs. non-type-Ia classification with host galaxy photo-z information. Without dropout the training loss continues to fall and the test loss rises. For 5 randomised runs, training for 200 epochs, we obtain a classification accuracy of $94.9 \pm 0.2$\%, AUC of $0.986 \pm 0.001$ and $F_1 = 0.64 \pm 0.01$. The corresponding type-Ia purity and completeness are $87.3 \pm 0.8\%$ and $91.4\pm 1.1\%$ respectively. A summary of results and comparisons can be found in Table~\ref{tab:05_SN1a}. The inclusion of host galaxy photo-$z$ marginally improves the classifier performance. The 1$\sigma$ errors quoted in the table are the result of 5 runs where the training data is randomly chosen (and so different) each time. Some random choice of the set of light-curves are more effective for training the network than others, but it is extremely difficult to optimise this.

To test the robustness of the time-grouping method we remove 10\% of the known filter values (and/or their errors) before grouping the data into a single vector and randomly augmenting the missing values. After training we find there is a small degradation in the results, i.e. for a training fraction of 0.5 using a deep 2-layer, unidirectional network with 16 hidden units, a dropout of 0.5 and including the photo-$z$ information the obtained results are very similar to the second line in Table~\ref{tab:05_SN1a}. This shows that a reduction in 10\% of the points is similar to the omission of the photo-$z$ data and therefore the data augmentation method is extremely robust.

\begin{table*}[t!]
\centering
\begin{tabular}{ c  c  c  c  c  c  c  c }\hline\hline
Method & Training size & AUC & Accuracy (\%) & $F_1$ & Purity (\%) & Completeness (\%) & Host $z$ \\\hline\hline
A & 10,660 & $0.986 \pm 0.001$ & $94.7 \pm 0.2$ & $0.64 \pm 0.01$ & $87.3 \pm 0.8$ & $91.4 \pm 1.1$ & True \\
A & 10,660 & $0.981 \pm 0.001$ & $93.6 \pm 0.3$ & $0.60 \pm 0.02$ & $87.4 \pm 1.7$ & $85.4 \pm 2.6$ & False \\
A & 5,330 & $0.975 \pm 0.003$ & $92.9 \pm 0.6$ & $0.57 \pm 0.03$ & $86.6\pm 2.0$ & $83.4 \pm 3.4$ & True \\
A & 5,330 & $0.973 \pm 0.002$ & $92.3 \pm 0.4$ & $0.55 \pm 0.02$ & $86.2 \pm 2.4$ & $80.8 \pm 3.8$ & False \\
B & 1,103 & $0.910 \pm 0.012$ & $85.9 \pm 0.9$ & $0.31 \pm 0.03$ & $72.4 \pm 0.4$ & $66.1 \pm 6.0$ & True \\
B & 1,103 & $0.901 \pm 0.016$ & $84.6 \pm 1.7$ & $0.28 \pm 0.05$ & $68.2 \pm 3.4$ & $66.3 \pm 5.5$ & False \\
C & $\sim$10,660 & - & - & 0.58 & 85 & 88 & True \\
C & $\sim$10,660 & - & - & 0.51 & 82 & 85 & False \\
C & 1,045 & - & - & 0.33 & 70 & 75 & True \\
C & 1,045 & - & - & 0.29 & 67 & 71 & False \\
D & $\sim$8,000 & - & - & 0.55 & - & - & True \\
D & $\sim$2,00$\underset{~}{0}$ & - & - & 0.45 & - & - & True \\
E & 1,103 & $0.94 \pm 0.03$ & - & - & - & - & True \\
E & 1,103 & $0.89 \pm 0.53$ & - & - & - & - & False \\
E & 1,103 & - & - & - & 90 & 85 & True\\
E & 1,103 & - & - & - & 87 & 90 & True\\
\hline\hline
F & 10,660 & $0.974 \pm 0.001$ & $90.4 \pm 0.3$ & - & $90.6 \pm 0.7$ & $86.5 \pm 0.7$ & True\\
F & 10,660 & $0.959 \pm 0.006$ & $88.5 \pm 1.1$ & - & $87.6 \pm 1.1$ & $85.9 \pm 4.1$ & False\\
G & 1,103 & $0.868 \pm 0.015$ & $78.1 \pm 0.9$ & - & $70.8 \pm 3.4$ & $70.6 \pm 4.1$ & True\\
G & 1,103 & $0.865 \pm 0.011$ & $78.0 \pm 1.2$ & - & $66.9 \pm 3.2$ & $74.5 \pm 4.2$ & False\\\hline\hline
A & 10,660 & $0.977 \pm 0.002$ & $93.1 \pm 0.4$ & $0.58 \pm 0.01$ & $88.0 \pm 1.1$ & $82.2 \pm 2.8$ & True\\
A & 10,660 & $0.970 \pm 0.001$ & $92.0 \pm 0.3$ & $0.53 \pm 0.01$ & $86.0 \pm 0.9$ & $79.5 \pm 2.2$ & False\\
B & 1,103 & $0.902 \pm 0.014$ & $85.2 \pm 1.2$ & $0.29 \pm 0.04$ & $71.5 \pm 1.6$ & $62.8 \pm 5.6$ & True\\
B & 1,103 & $0.860 \pm 0.017$ & $81.6 \pm 1.2$ & $0.21 \pm 0.02$ & $62.6 \pm 3.0$ & $57.6 \pm 2.7$ & False\\\hline\hline
A & 10,660 & $0.960 \pm 0.006$ & $87.9 \pm 0.9$ & - & $86.4 \pm 0.8$ & $84.4 \pm 3.5$ & True\\
A & 10,660 & $0.948 \pm 0.002$ & $86.8 \pm 0.3$ & - & $84.1 \pm 1.1$ & $83.7 \pm 1.4$ & False\\
B & 1,103 & $0.851 \pm 0.013$ & $76.8 \pm 1.3$ & - & $64.7 \pm 3.8$ & $71.0 \pm 4.1$ & True\\
B & 1,103 & $0.819 \pm 0.010$ & $74.2 \pm 1.0$ & - & $58.1 \pm 3.8$ & $73.6 \pm 6.6$ & False\\\hline\hline
\end{tabular}
\caption{\label{tab:05_SN1a} (Top section) Summary of results for  type-Ia vs. non-type-Ia classification with a training fraction of 0.5, 0.25 and 0.052 with comparisons to similar methods in~\cite{Karpenka:2012pm} and~\cite{Newling:2010bp}. (Second section) Summary of results for types-I,~-II~and~-III classification. (Third section) Summary of results for SPCC early-epoch challenge. (Bottom section) Summary of the results for the SPCC early-epoch challenge when classifying between types-I~,~-II~and~-III supernovae. The models used are A) Unidirectional LSTM, [16, 16] with 0.5 dropout, B) Unidirectional LSTM, [4] with 0.5 dropout, C) \cite{Karpenka:2012pm}, D) \cite{Newling:2010bp}, E) \cite{Lochner:2016hbn}~SALT2 fits averaged over machine learning architecture F) Bidirectional LSTM, [16, 16] with 0.5 dropout, G) Bidirectional LSTM, [4] with 0.5 dropout. Errors on results are the mean and standard deviation values from 5 randomised runs.}
\end{table*}

One advantage of our approach is that light curve data can be directly input to a {\em pre-trained} model to give very fast evaluation ($<1$s) of supernovae type. In the lower panel of Fig.~\ref{fig:lightcurve} we input the light curve, as a function of time, of a type-Ia supernovae (excluded from training) to the pre-trained 2-layer LSTM model discussed above. The classifier (type-Ia vs. non-type-Ia) is initially unsure of classification, with a type-Ia probability of around 0.5. The probability then decreases slightly, but rapidly increases near the peak of the light curve. The classifier has high confidence the supernovae is of type-Ia at around 60 days, and the final probability is excess of $99.5\%$.  This method could therefore be useful to give early indication of supernovae type in surveys.

We also test the same model using a training fraction of 0.25 (around 5000 supernovae), closer to the lower end of the number likely to be followed up for the LSST. After 5 randomised runs and training for 200 epochs we obtain an accuracy of $92.9 \pm 0.6\%$, AUC of $0.975 \pm 0.003$ and $F_1 = 0.57 \pm 0.03$. The corresponding type-Ia purity and completeness are $86.6\pm2.0\%$ and $83.4\pm3.4\%$ respectively. The $F_1$ metric has degraded by $\sim10\%$ for a reduction in data of $50\%$.

For $5.2\%$ of the representative SPCC data, the training dataset is so small that over-fitting is more severe. Using the same 2-layer LSTM network with 16 hidden units and dropout of 0.5 we find a notable increase in the test loss after $\sim 20$ epochs, but the accuracy and other metrics remain relatively constant ($F_1$ values of 0.35 to 0.4 were obtained). The reason for this apparent discrepancy is that the accuracy, say, simply takes the maximum value of the softmax output layer. For example, a 2-class problem with output probabilities [0.6, 0.4] and target [1, 0] has the same accuracy as one with output probabilities [0.8, 0.2]. The loss in the latter case would be lower however, and represents increased confidence of the network in its predictions. We therefore reject models with severe over-fitting and an increasing cross-entropy loss at the expense of metrics such as $F_1$, and decrease the model complexity. 

For a training fraction of  $5.2\%$ we find a single-layer LSTM network, with 4 hidden units, and dropout of 0.5 satisfies this criteria. For 5 randomised runs, training for 200 epochs, we obtain a classification accuracy of $85.9 \pm 0.9$\%, AUC of $0.910 \pm 0.012$ and $F_1 = 0.31 \pm 0.03$. The corresponding type-Ia purity and completeness are $72.4\pm0.4\%$ and $66.1\pm6.0\%$ respectively.

It is difficult to directly compare the results from the SPCC challenge in~\cite{Kessler:2010qj} with this work since the figure of merit is quoted as a function of redshift and a non-representative set of light-curves was originally used. In~\cite{Kessler:2010qj} the method of~\cite{2008AJ....135..348S} had the highest average $F_1$, with 79\% purity and 96\% accuracy. This is a, somewhat, confusing average as $F_1\sim 0.4$ at a redshift $z\sim0.1$ up to $F_1\sim1$ at $z\sim0.9$. Other methods performed similarly. 

It is better to consider comparison with other methods using post-SPCC data, for we obtain results which are competitive with previous approaches. The analyses by~\cite{Karpenka:2012pm} and~\cite{Newling:2010bp} are easier to compare. Along with~\cite{Lochner:2016hbn} these employ a two-step process, where features are first extracted by various methods before machine learning classification. The results obtained for similar sized training sets are comparable as can be seen in the top section of Table~\ref{tab:05_SN1a}. When using half the dataset to train on we get a higher $F_1$ value, $F_1=0.64$ compared to $F_1=0.58$ in~\cite{Karpenka:2012pm}. The value in~\cite{Newling:2010bp} is also similar given that the sample size is smaller. For a smaller sample training set of 5.2\% of all the data we again perform similarly to~\cite{Karpenka:2012pm} but under perform compare to~\cite{Newling:2010bp} taking into account the slightly larger sample size in the latter case. In~\cite{Lochner:2016hbn} using the SALT2 fits provided the best average AUC over a range of machine learning techniques. By imposing a purity of 90\% a completeness of 85\% was achieved while requiring a completeness of 90\% reveals a corresponding purity of 85\%.

In the second section of Table~\ref{tab:05_SN1a} the three class categorisation is shown. There is no available data for comparison of this problem, but compared to classification between type-Ia vs. non-type-Ia, bidirectional recurrent neural networks do well. The AUC and accuracy remain high, still above 90\% when the host-$z$ is included using a training fraction of 0.5. Using a smaller training fraction of 0.052, the results are worsened similar to the two class categorisation in the top section of Table~\ref{tab:05_SN1a}.

The third section of Table~\ref{tab:05_SN1a} shows the results of the early-epoch challenge from SPCC. Here only the data before the night of the sixth observation with $S/N>4$ for each light-curve can be used - a great reduction from the use of the full light-curve. We do surprisingly well in this case obtaining an accuracy of $93.1\pm0.4\%$, AUC of $0.977\pm0.002$ and an $F_1=0.58\pm 0.01$ with a training fraction of 0.5 and including host-$z$. These values are not far from those obtained using the whole light-curve and are equivalent to the full results of~\cite{Karpenka:2012pm}. The results are not as good with a training fraction of 0.052, but still comparable to our results using the whole light-curve. The network trained on the partial light-curves does better than suggested from feeding the early-epoch light-curve through a network trained on the full sequence. This is due to the later parts of the light-curve influencing the weights of the network whilst training. Training on only the initial part of the light-curve optimises the network weights such that early sequence features have more effect, resulting in better accuracy, AUC and $F_1$ values than expected.

Finally, the bottom section of Table~\ref{tab:05_SN1a} has the results of the three class categorisation when using the early-epoch data. The results are similar to the difference between the full light-curve and early-epoch data SN1a categorisation when comparing with the full light-curve 123 categorisation. It should be noted that the bidirectional network used for the 123 categorisation using the full light-curve revealed sizeable over-fitting when using the early-epoch data and so a unidirectional network was used instead.

\section{Conclusions}

We have presented a new method for performing photometric classification of supernovae. Machine learning methodology has previously been applied to SPCC classification~\cite{Newling:2010bp, Karpenka:2012pm, Lochner:2016hbn}. Instead of performing feature extraction before classification, our approach uses the light-curves directly as inputs to a recurrent neural network, which is able to learn information from the sequence of observations.

Although we have trained the network on the cross-entropy loss and not the $F_1$ score, for the same sized dataset of $\sim10^3 (10^4)$ supernovae (including host galaxy photo-$z$), \cite{Karpenka:2012pm} obtained $F_1$ values of 0.33 (0.58), and \cite{Newling:2010bp} values of 0.42 (0.57), compared to our  0.31 (0.64). Recurrent neural networks  therefore compare well with other methods when a larger training  set is available. The performance isn't  quite as good with a smaller training set, possibly due to the network having to learn from no prior information about (noisy) light curves. The current state-of-the-art for a small training set ($\sim10^3$ supernovae) comes from a combination of SALT2 (Spectral Adaptive Light curve Template 2) template fits and boosted decision trees~\cite{Lochner:2016hbn}. It would be interesting to check how how deep learning compares to this with a larger training set.

As well as finding competitive results for the final metrics, we have shown that it is possible to give fast, early evaluation of supernovae type using pre-trained models. This is possible since the light curve can be fed to the model directly without needing any feature extraction. 

Most interestingly, we have found that training a network only on the early epoch light-curve data results in a better early-time predictor than using a network trained on entire light-curve data. Our results using only the early-epoch data are close to those using the entire light-curve data for both SN1a and 123 categorisation with both large and small training fractions.

There are several possibilities for future work. One of the advantages of recurrent neural networks is that inputs are agnostic, so the impact of any additional inputs could be explored. It would be possible, for example, to even pass the raw images in each filter though a convolutional network and use those as inputs. We have considered a representative training sample, but spectroscopic follow up surveys may be biased. The performance of the network could be measured against selection bias, and the results used to inform the best follow up strategy.  Further work could also be performed to optimise the early detection probability of the network.  Finally, to improve performance in the small data regime one can use {\em transfer learning.} Here, a more complex network is pre-trained on simulations or existing data from other surveys, then the weights of the network are fine-tuned on the new, smaller dataset. The simulated SPCC data used in this work are based on the DES instrument, and we are applying transfer learning to real DES data for publication in future work. 

\section*{Acknowledgements}

We appreciate helpful conversations with Steven Bamford, Simon Dye, Mark Sullivan and Michael Wood-Vasey, and Natasha Karpenka, Richard Kessler and Michelle Lochner for help with data acquisition. T.C. is supported by a STFC studentship, and A.M. is supported by a Royal Society University Research Fellowship.

\end{document}